\documentclass{appolb}
\usepackage{amsfonts}
\usepackage{amsmath}
\usepackage{graphicx}
\usepackage{nicefrac}
\usepackage{hyperref}
\usepackage{cite}
\begin{document}
% \eqsec  % uncomment this line to get equations numbered by (sec.num)
\title{Complex Langevin in Lattice QCD: dynamic stabilisation and the phase diagram%
%
% you can use '\\' to break lines
}
\author{Gert Aarts, Felipe Attanasio\thanks{Presented at the Workshop ``Excited QCD 2016'', Portugal 6-11 March 2016}, Benjamin J\"ager
\address{Department of Physics, College of Science, Swansea University, United Kindgom}
\\~\\
{D\'enes Sexty
}
\address{Department of Physics, Bergische Universit\"at Wuppertal, Wuppertal, Germany}
}
\maketitle
\begin{abstract}
Complex Langevin simulations provide an alternative to sample path integrals with complex weights and therefore are suited to determine the phase diagram of QCD from first principles.
We use our proposed method of Dynamic Stabilisation (DS) to ensure improved convergence to the right limit and present new systematic tests of this technique.
%We also show results on QCD in the limit of heavy quarks and preliminary results with fully dynamical staggered quarks.
We also show results on QCD in the limit of heavy quarks and an analysis of DS compared to known results from reweighting.
\end{abstract}
  
\section{Introduction}
The QCD phase diagram is subject of active research both theoretically and experimentally.
On the theoretical side the phase structure of QCD has implications for the understanding of the early and modern universe, where the quark-gluon plasma and compact astrophysical objects still pose unanswered questions.
For experimentalists, it of course serves as guide for current and future experiments and facilities.

Studies of the phase diagram at non-zero chemical potential $\mu$ are extremely difficult due to a complex statistical weight, which results in what is known as the \textit{sign problem}.
In QCD this behaviour is caused by the quark determinant, after the quark fields have been integrated out
\begin{align}
	\left[ \det M(U,\mu) \right]^\ast = \det M(U,-\mu^\ast)\,,
\end{align}
where $U$ generically represents the gauge links.
This leads to an exponentially small overlap between the full theory and the ``phase quenched'' version which can be simulated using Monte Carlo techniques.
A review of different methods for the QCD phase diagram can be found in \cite{deForcrand:2010ys}.

\section{Complex Langevin equation}
The complex Langevin method is based on stochastic quantisation \cite{Parisi:1980ys}.
In this framework the dynamical variables are evolved along a fictitious time direction $\theta$ according to a Langevin equation.
On the lattice for an SU($3$) gauge theory the equation for the gauge links $U$ reads \cite{Damgaard:1987rr}
\begin{equation}
U_{x\mu}(\theta + \varepsilon) = \exp\left[ i \lambda^a X^a_{x\mu} \right] U_{x\mu}(\theta)\,,\quad X^a_{x\mu} = -\varepsilon D^a_{x\mu} S[U] + \sqrt{\varepsilon} \eta^a_{x\mu}\,,
\end{equation}
where $\lambda^a$ are the Gell-Mann matrices normalised to $\Tr[\lambda^a \lambda^b] = 2 \delta^{ab}$, $\varepsilon$ is the Langevin stepsize chosen adaptively \cite{Aarts:2009dg}, $\eta^a_{x\mu}$ are white noise fields satisfying $\langle \eta^a_{x\mu} \eta^b_{y\nu} \rangle = 2 \delta^{ab} \delta_{xy} \delta_{\mu\nu}$, $S[U]$ is the action, and the gauge group derivative $D^a_{x\mu}$ is defined as
\begin{equation}
D^a_{x\mu} f(U) = \left.\frac{\partial}{\partial \alpha} f(e^{i \alpha \lambda^a} U_{x\mu})\right|_{\alpha = 0}\,.
\end{equation}

To deal with the sign problem we allow the gauge links to be non-unitary \cite{Klauder:1983zm, Parisi:1984cs, Aarts:2009uq}, which amounts to extending the group SU($3$) to SL($3,\mathbb{C}$).
The latter is a non-compact manifold and therefore large excursions into the imaginary directions may occur during the simulation.
%These are known to cause convergence to a wrong limit [].

As a way to keep track of such excursions and prevent them from becoming large -- by using gauge transformations between Langevin updates -- we measure the distance from SU($3$),
\begin{equation}
d = \frac{1}{N_s^3 N_\tau} \sum_{x,\mu} \Tr \left[ U_{x\mu} U^\dagger_{x\mu} - \mathbf{1} \right]^2 \geq 0 \,,
\end{equation}
where $N_s$ and $N_\tau$ are the lattice extents in the spatial and temporal directions, respectively.
These transformations, known as gauge cooling \cite{Seiler:2012wz}, are constructed to minimise $d$ and are defined as
\begin{equation}
U_{x\mu} \to \Lambda_x U_{x\mu} \Lambda^{-1}_{x+\mu}\,, \quad \Lambda_x = \exp\left[ -\varepsilon \alpha \lambda^a f^a_x \right]\,,
\end{equation}
where
\begin{equation}
f^a_x = 2 \Tr \left[ \lambda^a \left( U_{x\mu} U^\dagger_{x\mu} - U^\dagger_{x-\mu,\mu} U_{x-\mu,\mu} \right) \right]\,.
\end{equation}
The parameter $\alpha$ and the number of cooling steps are chosen adaptively in order maximise the method's efficiency \cite{Aarts:2013uxa}.
A study of this method applied to one- and two-dimensional QCD can be found in \cite{Bloch:2015coa}.

\section{QCD with heavy quarks}
In the heavy quarks approximation \cite{Bender:1992gn, Aarts:2008rr} the quarks evolve only in the Euclidean time and the fermion determinant decouples as a product over spatial points.
This corresponds to the leading order in a spatial hopping expansion,
\begin{equation}
	\det M(U, \mu) = \prod_{\vec{x}} \left\{ \det \left[ 1 + (2\kappa e^{\mu})^{N_\tau} \mathcal{P}_{\vec{x}} \right] \det \left[ 1 + (2\kappa e^{-\mu})^{N_\tau} \mathcal{P}^{-1}_{\vec{x}} \right] \right\}^2\,,
\end{equation}
containing the hopping parameter $\kappa$, the Polyakov loop and its inverse
\begin{equation}
		\mathcal{P}_{\vec{x}} = \prod^{N_\tau-1}_{\tau = 0} U_4(\vec{x})\,,\quad \mathcal{P}^{-1}_{\vec{x}} = \prod^0_{\tau=N_\tau-1} U^{-1}_4(\vec{x})\,.
\end{equation}
This model exhibits a sign problem and a transition to a high density phase.
At zero temperature this is expected to happen at $\mu = \mu^0_c \equiv -\ln(2\kappa)$.

Comparisons with the hopping expansion to all orders \cite{Aarts:2014bwa} as well as multi-parameter reweighting \cite{Fodor:2015doa} have been investigated recently.
For a study of a strong coupling expansion ($\beta \to 0$) combined with hopping parameter expansion see \cite{deForcrand:2014tha, Glesaaen:2015vtp}.
Effective model studies for heavy quarks QCD include \cite{Scior:2016fso, Ejiri:2015dzh}.
Complex Langevin investigations can be found in \cite{Aarts:2014kja, Aarts:2014fsa, Aarts:2015yba, Aarts:2015hnb, Aarts:2015yuz, Aarts:2016qrv}.

\begin{figure}[htb]
\centerline{%
\includegraphics[width=8.5cm]{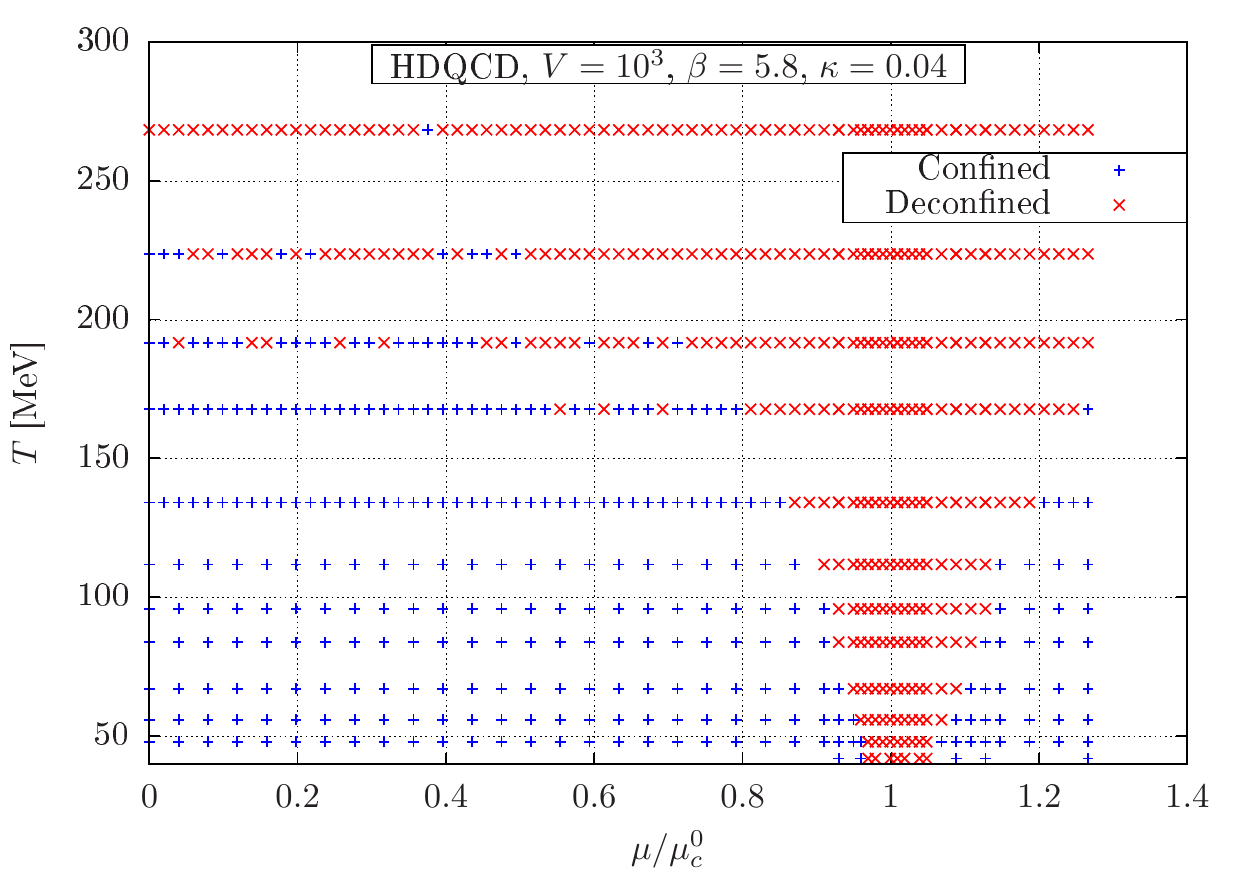}}
\caption{Phase boundary of the HDQCD model. The phase boundary was defined as the region where the Binder cumulant transitions from zero to non-zero values within statistical precision. At $\mu/\mu^0_c>1$ we see a similar pattern, which is an effect from particle-hole symmetry \cite{Rindlisbacher:2015pea}.}
\label{Fig.HD.GC.Binder.binary.capped}
\end{figure}

The expectation value of the Polyakov loop is known to be approximately $0$ in a confined phase and non-zero otherwise.
We looked at its Binder cumulant \cite{Binder:1981sa} to map the boundary between these phases.
The resulting plot is shown in figure \ref{Fig.HD.GC.Binder.binary.capped}, where a lattice spacing of $a \sim 0.15 \text{ fm}$, determined using the Wilson flow \cite{Borsanyi:2012zs}, has been used to convert the temperature to physical units.
Our analysis included only data with small unitarity norm ($d \leq 0.03$) to allow for reliable predictions, as will be explained later.

\section{Dynamic Stabilisation}
There are situations where gauge cooling is not sufficient to keep the system from going too far into SL($3,\mathbb{C}$) and when the distance becomes $\mathcal{O}(0.1)$ observables seem to converge to a wrong limit \cite{Aarts:2013uxa}.

We have developed a new technique called dynamic stabilisation, which consists of adding a SU($3$) -- but not SL($3,\mathbb{C}$) -- gauge invariant force to the Langevin drift that is trivial in the continuum limit and grows with $d$.
It is given by
\begin{equation}
		M^a_x[U] = i b^a_x \left( b^c_x b^c_x \right)^3 \,, \quad b^a_x = \sum_\nu \Tr \left[ \lambda^a U_{x\nu} U^\dagger_{x\nu} \right]\,.
\end{equation}
To better benefit from this technique we have also extended our Langevin equation to second order in the Langevin stepsize \cite{Fukugita:1986tg}
\begin{align}
	U_{x\mu}(\theta+\nicefrac{\varepsilon}{2}) &= \exp\left[i \lambda^a X^a_{x\mu}\right] U_{x\mu}(\theta)\,,\\
	U_{x\mu}(\theta+\varepsilon) &= \exp\left[i \lambda^a \gamma \left(X^{\prime a}_{x\mu} + X^a_{x\mu}\right)\right] U_{x\mu}(\theta)\,,
\end{align}
where the new drifts read
\begin{align}
		X^a_{x\mu} &= -\varepsilon D^a_{x\mu}S\left[U(\theta)\right] + i \varepsilon \alpha_{DS} M^a_x[U(\theta)] + \sqrt{\varepsilon} \, \eta^a_{x\mu}(\theta)\,,\\
				X^{\prime a}_{x\mu} &= -\varepsilon D^a_{x\mu}S\left[U(\theta+\nicefrac{\varepsilon}{2})\right] + i \varepsilon \alpha_{DS} M^a_x[U(\theta + \nicefrac{\varepsilon}{2})] + \sqrt{\varepsilon} \, \eta^a_{x\mu}(\theta)\,,
\end{align}
with $\gamma = \nicefrac{1}{2}\left(1 + \varepsilon C_A/6\right)$, $\alpha_{DS}$ a real parameter, the noise is $\langle \eta^a_{x\mu} \eta^b_{y\nu} \rangle = 2 (1 - \varepsilon C_A/2) \delta^{ab} \delta_{xy} \delta_{\mu\nu}$ and $C_A=3$ is the Casimir invariant in the adjoint representation.

\begin{figure}[htb]
\centerline{%
\includegraphics[width=8.0cm]{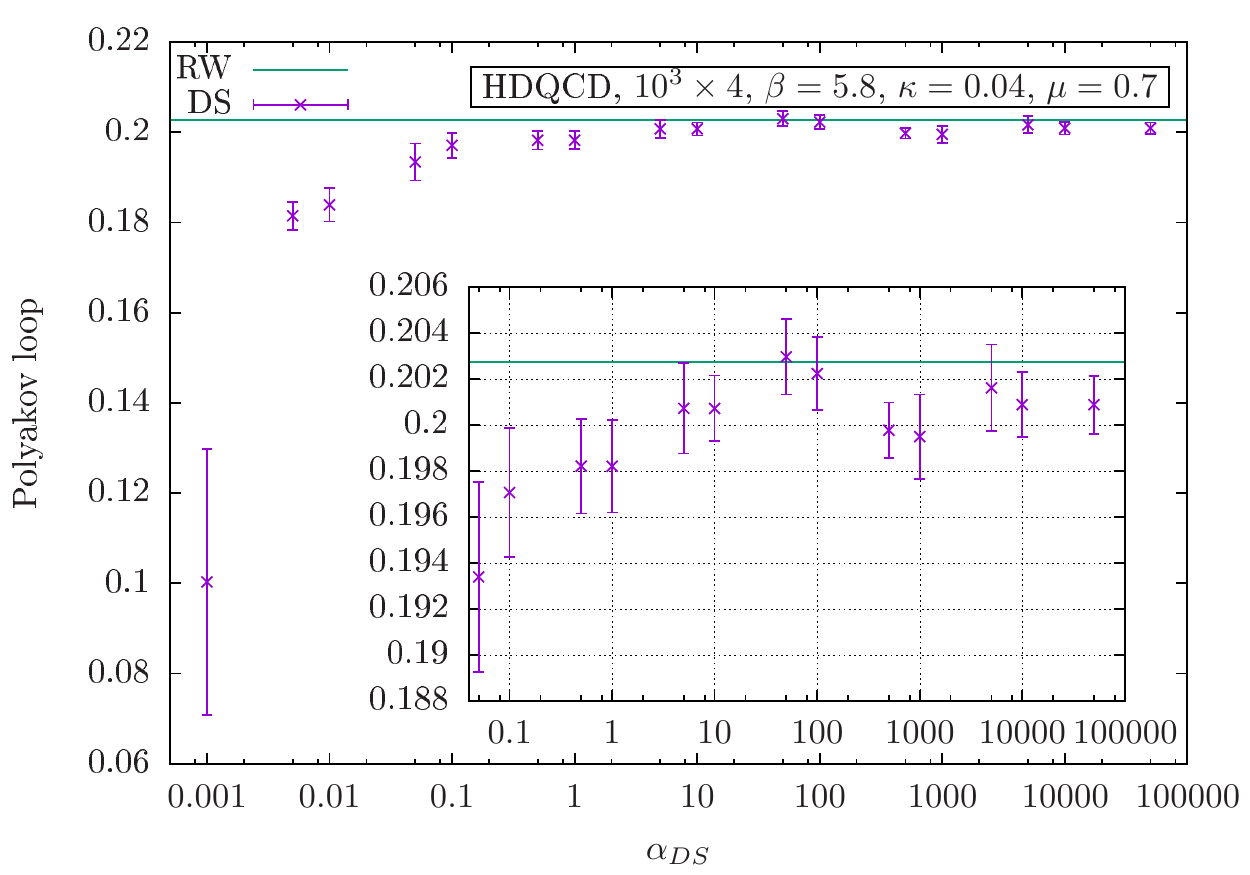}}
\vspace{-3mm}
\caption{Polyakov loop in the HDQCD model for different values of $\alpha_{DS}$. The inset shows a zoom into the region of agreement.}
\label{Fig.HD.DS.alpha.scan}
\vspace{-3mm}
\end{figure}

\iffalse
The new Langevin drift then reads, with $\alpha$ being a control parameter,
\begin{equation}
	X^a_{x\mu} = \varepsilon D^a_{x\mu} S[U] + \varepsilon \alpha M^a_x + \sqrt{\varepsilon} \eta^a_{x\mu}\,.
\end{equation}

A comparison between gauge cooling and dynamic stabilisation with a known result from reweighting is shown in figure \ref{Fig.HD.DS.GC.RW}, while figure \ref{Fig.HD.DS.alpha.scan} shows the same physical setup with different $\alpha$ compared with reweighting.
\vspace{-3mm}
\begin{figure}[htb]
\centerline{%
\includegraphics[width=10.5cm]{./images/polyakov.pdf}}
\vspace{-3mm}
\caption{Comparison of HDQCD model using dynamic stabilisation, gauge cooling and reweighting.}
\label{Fig.HD.DS.GC.RW}
\end{figure}
\fi
A comparison between dynamic stabilisation combined with one gauge cooling step with a known result from reweighting, $\langle P \rangle = 0.202717(66)$, is shown in figure \ref{Fig.HD.DS.alpha.scan} for different $\alpha_{DS}$'s in a lattice of volume $10^3 \times 4$, $\beta=5.8$, $\kappa=0.04$ and $\mu=0.7$.
Agreement with reweighting results is seen for $\alpha_{DS}$ sufficiently large.
For this setup gauge cooling gives $\langle P \rangle = 0.0027(55)$ if the points with unitarity norm larger than $0.03$ are included.
%\vspace{-4mm}

%\vspace{-1.1cm}
\section{Conclusions and Outlook}
\vspace{-0.15cm}
We have shown that the complex Langevin equation method allows simulation of theories that exhibit the sign problem, provided that large excursions into the complex directions are suppressed.
The gauge cooling technique provided the first consistent way of achieving that and allowed us to map the phase boundary in the limit of heavy quarks.

Our method of dynamic stabilisation, combined with gauge cooling, provides greater control on how far from SU($3$) the system can go.
It has been shown to agree with known results in the heavy quarks approximation.
Further tests with other physical parameters are needed to ensure its reliability.

The next step is to apply these techniques to QCD with fully dynamical quarks \cite{Sexty:2013ica}.

\vspace{-0.6cm}
\section{Acknowledgements}
\vspace{-0.15cm}
We are grateful for the computing resources made
available by HPC Wales.
We acknowledge
the STFC grants ST/L000369/1,                                                         % Swansea theory group consolidated grant
ST/K000411/1,  ST/H008845/1, ST/K005804/1 and ST/K005790/1,     % DiRAC grants
the STFC DiRAC HPC Facility (www.dirac.ac.uk),
STFC grant ST/L000369/1, the
Royal Society and the Wolfson Foundation. FA is grateful for the support
through the Brazilian government programme ``Science without Borders'' under scholarship number
BEX 9463/13-5.

\end{document}